\documentclass[12pt]{article}
\begin{document}
\title{\bf The Canonical Form of the Rabi Hamiltonian}
\author{M. Szopa\thanks{On leave of the Institute of Physics, 
University of Silesia, Uniwersytecka 4, 40-007 Katowice, 
Poland}, G. Mys and A. Ceulemans\\
Department of Quantum Chemistry, University of Leuven,\\
Celestijnenlaan 200F, B-3001 Leuven, Belgium.}

\baselineskip=17pt

\date{}
\maketitle

\begin{abstract}
The Rabi Hamiltonian, describing the coupling of a two-level system to a 
single quantized boson mode, is studied in the Bargmann-Fock 
representation. The corresponding system of differential equations is 
transformed into a canonical form in which all regular singularities 
between zero and infinity have been removed. The canonical or 
Birkhoff-transformed equations give rise to a two-dimensional eigenvalue 
problem, involving the energy and a transformational parameter which 
affects the coupling strength. The known isolated exact solutions of the 
Rabi Hamiltonian are found to correspond to the uncoupled form of the 
canonical system.

PACS numbers: 03.65.Ge, 33.20.Wr.
\end{abstract}

                   \section*{I. Introduction}

In the study of dynamical problems the harmonic oscillator
occupies a prominent place, as a prototype of the fundamental
unitary symmetry group. The spell of group theory also extends 
to anharmonic oscillators, which have recently been exposed 
as mere q-deformations of
the unitary group algebra \cite{kibl}. In contrast dynamical
problems which arise in the study of boson-fermion interactions
are more reluctant to reveal their hidden symmetries.
Such problems are more involved since the actual states of the
system are of composite nature with a boson and a fermion part.
This results in the appearance of singularities in the
corresponding eigenvalue problem. Examples include the
Jahn-Teller Hamiltonian in molecular physics, and the Rabi
Hamiltonian in nuclear physics and quantum optics. 

A rather peculiar feature of these systems is the emergence of
isolated exact solutions, with eigenvalues that correspond to
simple expressions in rational numbers. Such were obtained - in
a rather heuristic way - by Judd \cite{judd} for the case of
the $E\otimes e$ Jahn-Teller Hamiltonian, and by Ku\'s
\cite{kus} for the spectrum of a two-level atom coupled to a
single quantized mode. It was suggested that the exact solutions
probably hint at some dynamical symmetry group - but so far no
progress was reported in this direction. 

In the present paper we analyze the one mode Rabi problem in a more
rigorous way.  First - following \cite{kus} and \cite{reik} -
the dynamical problem is defined in the Bargmann-Fock Hilbert
space of entire functions. The resulting system of differential
equations is then put into canonical form using a theorem due 
to Birkhoff. Under this transformation the Ku\'s exact solutions are
found to be mapped onto the levels of a displaced harmonic
oscillator. In this way hidden symmetry appears.

  \section*{II. The Bargmann-Fock Representation of a Hilbert Space}

Before turning to the actual results we briefly review basic
information about the Hilbert space of entire functions
introduced by Fock \cite{fock} and Bargmann \cite{barg}.  Let
${\cal F} _{n}$, for n integral, be the set of entire analytic
functions $f(z)$, where $z=(z_{1},...,z_{n})$ and $z_{k}\in
\bf{C}$ are complex numbers, $k=1,...,n$. Because $f(z)$ is
entire it has an everywhere converging power series
    \begin{equation}
    f(z) = \sum_{k_{1},...,k_{n}} \alpha_{k_{1},...,k_{n}} 
    z_{1}^{k_{1}}\cdot ...\cdot z_{n}^{k_{n}},   \label{fodz}
    \end{equation} 
where summation extends over the whole set of
non-negative integers $k_{1},...,k_{n}$ and $\alpha_{k_{1},...,k_{n}}$
are complex coefficients. We define an inner product of two elements 
$f$ and $g$ of ${\cal F} _{n}$ by
    \begin{equation}
    (f,g) = \int \overline{f(z)} \cdot g(z) d\mu_{n}(z),   
     \label{ilsk}     
    \end{equation} 
where $\overline{f(z)} \cdot g(z)$ is a usual scalar product
in ${\bf C}^{n}$,
    \begin{equation}
     d\mu_{n}(z) = \frac{1}{\pi^{n}} \exp(-\overline{z} 
     \cdot z) \prod_{k=1}^{n} dx_{k} dy_{k}, \hspace{1.3cm}
     z_{k} = x_{k} + iy_{k}   \label{miar}
    \end{equation}
and the integration extends over the whole space ${\bf C}^{n}$.
The Bargmann-Fock space ${\cal F} _{n}$  is a set of all entire 
functions (\ref{fodz}), which have a finite norm 
$(f,f)< \infty$. This is equivalent to the requirement that 
    \begin{equation}
    |f(z)| \leq c \exp(\frac{1}{2}\gamma \overline{z}\cdot 
     z),     \label{waru}
    \end{equation}
where $c$ and $\gamma$ are positive constants with $\gamma < 1$.
The Bargmann-Fock space  ${\cal F} _{n}$ with 
the inner product defined by (\ref{ilsk}) is a Hilbert space.

Let us now consider two operators in ${\cal F} _{n}$: 
multiplication by $z_{k}$ and differentiation 
$\frac{d}{dz_{k}}$. Since the functions $f(z)$ of  ${\cal F} 
_{n}$ are analytic, $z_{k}f$ and $\frac{d}{dz_{k}} f$ always 
exist. The operators satisfy the commutation rules
    \begin{equation}
    \left[z_{k},z_{l}\right] = 0,\hspace{1cm} 
    \left[\frac{d}{dz_{k}}, \frac{d}{dz_{l}}\right] = 0,\hspace{1cm}
    \left[\frac{d}{dz_{k}}, z_{l}\right] = \delta_{kl}.
    \label{komm}
    \end{equation}
Furthermore, with respect to the inner product (\ref{ilsk}) 
$z_{k}$ and $\frac{d}{dz_{k}}$ are Hermitian conjugate
    \begin{equation}
    \left(z_{k} f, g\right) =  \left(f, \frac{d}{dz_{k}} g\right)
    \label{herm}
    \end{equation}
whenever $z_{k}f$ and $\frac{d}{dz_{k}} f$ belong to ${\cal 
F} _{n}$. 

On the other hand relations (\ref{komm}) and (\ref{herm}) 
are well known algebraic relations defining Hermitian 
conjugate annihilation 
$a_{k}$ and creation $a_{k}^{+}$ operators of boson fields 
in second quantization,
    \begin{equation}
    \left[a_{k}^{+}, a_{l}^{+} \right] = 0,\hspace{1cm}
    \left[a_{k}, a_{l} \right] = 0,\hspace{1cm}
    \left[a_{k}, a_{l}^{+} \right] = \delta_{kl}.
    \label{koma}
    \end{equation}
We conclude therefore that within ${\cal F} _{n}$ the 
annihilation operator $a_{k}$ is represented by the operation 
$\frac{d}{dz_{k}}$, and the creation operator $a_{k}^{+}$ 
corresponds to the multiplication by $z_{k}$. 

As an instructive example let us take a set of n identical uncoupled 
harmonic oscillators which are described up to a constant by the 
Hamiltonian $H = \sum_{l=1}^{n} a_{l}^{+} a_{l}$. The corresponding 
operator in the Bargmann-Fock space (denoted by ${\cal H}$) is 
${\cal H}\ = \sum_{l=1}^{n} 
z_{l} \frac{d}{dz_{l}}$ and the corresponding eigenproblem is the 
system of differential equations
    \begin{equation}
      \sum_{l=1}^{n} z_{l} \frac{d}{dz_{l}} f_{k}(z) = E_{k} 
      f_{k}(z).    \label{eigo}
    \end{equation}
It is easily verified that the eigenfuctions in this case are 
functions of the type (\ref{fodz}), which are homogeneous 
polynomials of the order $k$ i.e. in (\ref{fodz}) the 
sum is over $k_{1}+...+k_{n} = k$ and the corresponding eigenvalues 
are $E_{k}=k$.

 \section*{III. The Transformation to a Canonical Form}

In the Bargmann-Fock representation of a Hilbert space quantum
mechanical equations for bosons interacting with a manifold of
fermion states are represented by a system of linear
differential equations in the complex domain \cite{kus,reik}.
The physical solutions (\ref{fodz}) of such a system must belong
to ${\cal F} _{n}$ i.e. be entire and obey condition
(\ref{waru}). In practice the solution of this equation is
complicated due to the occurrence of finite regular
singularities. 

In this section we describe a transformation - due to Birkhoff
\cite{birk} - which allows, for a system of linear equations of
the first order in one variable, to find a canonical form. To be
in the canonical form the system must be transformed in such a
way that all its finite singularities reduce to only one
singularity at zero, while at the same time preserving the order
of the singularity at infinity. Consequently, the transformed
system is more likely to be exactly solvable. 

The system of m linear differential equations of the first order
has a general form
    \begin{equation}
    \frac{df_r}{dz} = \sum_{s=1}^{m} p_{rs}(z) f_{s},
    \hspace{2cm} r=1,...,m \label{insy}
    \end{equation} 
and we assume that $p_{rs}(z)$ are analytic
functions of a complex variable apart from a finite number of
regular singularities (even at infinity). Practically it means,
that outside the circle $|z|=R$, which includes all the finite
singular points, the coefficients may be expanded in a Laurent
series
    \begin{equation}
    p_{rs}(z) = \sum_{k=-\infty}^{q} p_{rs}^{(k)} z^{k} ,
    \hspace{2cm} p_{rs}^{(k)}\in \bf{C}, \label{expp}
    \end{equation}
where $q\geq-1$ and $q+1$ is termed the {\em rank} of the
singular point at infinity. In general the system (\ref{insy})
can have $m$ independent sets of solutions $f_{s}^{(t)}(z)$,
where $s=1,...,m$ denotes different solutions within a set
$t=1,...,m$. Now we assume a linear transformation of the form
    \begin{equation}
    f_{r}(z) = \sum_{s=1}^{m} a_{rs}(z) F_{s}(z)   \label{tran}
    \end{equation}
where the coefficients $a_{rs}(z)$ are analytic at infinity and 
reduce at infinity to a unit matrix
    \begin{equation}
    a_{rs}(z) = \sum_{k=0}^{\infty} \frac{a_{rs}^{(k)}}{z^{k}} ,
    \hspace{1.5cm} a_{rs}^{(k)}\in {\bf C}, \hspace{0.5cm} 
    a_{rs}^{(0)}=\delta_{rs}.  
    \label{mata}
    \end{equation}
In a sense this $\{a_{rs}(z)\}$ matrix could be said to contain
all the finite singularities of the initial system. Under this
transformation the original system (\ref{insy}) turns into a
system of a slightly different form
    \begin{equation}
    z \frac{dF_{r}(z)}{dz} = \sum_{s=1}^{m} P_{rs}(z) F_{s}(z),
    \hspace{2cm} r=1,...,m. \label{trsy}
    \end{equation}
The coefficients of the transformed system are given by the equation
   \begin{equation}
\{P_{rs}(z)\} =z \left(\left\{a_{rk}^{-1}(z)\right\} \left\{p_{kl}(z)
   \right\} \left\{a_{ls}(z)\right\} - \left\{a_{rk}^{-1}(z)\right\} 
     \left\{\frac{d}{dz} a_{ks}(z)\right\} \right),   \label{eqfP}
    \end{equation}
where $\{a_{rk}^{-1}(z)\}$ is the matrix inverse to the matrix 
$\{a_{rk}(z)\}$ and $\{a_{rk}(z)\} \{p_{ks}(z)\}$ denotes the matrix 
multiplication.
Now we are in the position to formulate the Birkhoff theorem - the 
crucial result in our analysis. 

The Birkhoff theorem states that {\em for every system of the
type (\ref{insy}) there exists a transformation matrix
(\ref{mata}) such that the coefficients $P_{rs}(z)$ of the
transformed system (\ref{trsy}) are polynomials of a degree not
exceeding $q+1$} \cite{birk}. 

There are several properties of the above transformation which 
should be stated here.\\
(i) All finite singular points of the initial system (\ref{insy}) 
coalesce to only one singularity at zero (because $P_{rs}(z)$ are 
polynomials). This is the most 
significant property of the transformation and because of it the 
system (\ref{trsy}) may be termed the {\em canonical} form of 
the system (\ref{insy}).\\
(ii) The ranks of the singular points at infinity for both systems 
(\ref{insy}) and (\ref{trsy}) are equal.\\
(iii) If a given set of solutions  $f_{s}^{(t)}(z)$, $s=1,...,m$, 
belongs to the Bargmann-Fock 
space  ${\cal F}_{1}$, then the corresponding transformed solutions  
$F_{s}^{(t)}(z)$, which are found due to the Birkhoff theorem, also 
belong to this space. This can easily by shown by adapting the 
treatment by Birkhoff to the case of entire functions.

The final property allows to reject all the solutions of the
transformed system which do not belong to the Bargmann-Fock
space as being non physical. However the inverse of this
property is not automatically true, but requires a proper choice
of the transformation matrix. We will come back to this point
when discussing the actual solutions of the system under
investigation. 

A useful test to check if the solutions of the transformed system 
can be entire is given by the indicial equation of the transformed 
system \cite{ince}
    \begin{equation}
    det\{c_{rs} - \rho \delta_{rs}\} = 0    \label{indi}
      \end{equation}
where $c_{rs} = P_{rs}(0)$, $r,s=1,...,m$. The solutions of 
(\ref{trsy}) 
depend on the roots $\rho_{1},...,\rho_{m}$ of the indicial 
equation. There are several rules connected with this equation 
which indicate the possibility of existence of entire solutions and 
their degeneracy.\\
$D_{0}$. If none of the roots is a non-negative integer, the equation 
(\ref{trsy}) has no entire solutions (because they are not 
analytic in the origin).\\
$D_{1}$. If one of the roots $\rho_t$ is a non-negative integer and 
the remaining roots are either non-integers or are equal to $\rho_t$, 
then there exists exactly one, up to linear dependence, set of 
analytic solutions of (\ref{trsy}) and it is of the form
    \begin{equation}
    F_{s}^{(t)}(z) = z^{\rho_{t}} u_{s}(z), \hspace{1.5cm} 
    s=1,...,m \label{fors}
    \end{equation}
where the $u_{s}(z)$ are analytic and $u_{s}(0)\neq 0$.  Functions
(\ref{fors}) are entire provided their radii of convergence are
infinite.\\ $D_{2}$. In the remaining cases if two or more roots
of the indicial equation are integers (at least one of them is
non-negative), there exists at least one analytic set of
solutions of the form (\ref{fors}), where $\rho_{t}$ is the
maximal integral root of (\ref{indi}). The remaining solutions
corresponding to other integral roots are usually singular in
the origin, but in exceptional cases may also be analytic. 

It should be pointed out that the Birkhoff theorem is an
existence theorem, which as such does not provide the actual
form of the canonical equation nor the transformation matrix. In
practice at least the canonical equation can usually be found
relatively easy by the use of (\ref{eqfP}), keeping in mind that
the $P_{rs}(z)$ are polynomials of a given degree. To this aim
we invert the equation (\ref{eqfP}) and express it in terms of
the relevant expansion coefficients. This yields, for every
integer $l\geq 0$, a system of $m^{2}$ equations
    \begin{equation}
     \sum_{i=0}^{l} \left(\left\{a_{rk}^{(l-i)}\right\} 
     \left\{P_{ks}^{(q+1-i)}\right\} - 
     \left\{p_{rk}^{(q-i)}\right\} \left\{a_{ks}^{(l-i)}\right\} 
     \right) = (l-q-1) \left\{a_{rs}^{(l-q-1)}\right\},
     \label{eqfa}
    \end{equation}
where expansion coefficients for $\{p_{rk}(z)\}$, $\{a_{rk}(z)\}$ 
and $\{P_{rs}(z)\}$ are defined by (\ref{expp}), (\ref{mata}) and 
$P_{rs}(z) = \sum_{k=0}^{q+1} P_{rs}^{(k)} z^{k}$ respectively. 
We assume also that $\{a_{rk}^{(i)}\}$ and 
$\{P_{rk}^{(i)}\}$ with negative indices $i$ are zero matrices.
 
This formula can now be used to find the coefficients
$P_{rs}(z)$ of the transformed system. In this case we only need
the equations corresponding to $l=0,...,q+1$. Here the trivial
case with $l=0$ immediately yields: 
    \begin{equation}
     \{P_{rs}^{(q+1)}\} =  \{p_{rs}^{(q)}\}.
     \label{zoep}
    \end{equation}
For higher $l$, $1\leq l \leq q+1$ the resulting expansion
coefficients in the transformed system may also depend on the
$a_{ks}^{(1)},...,a_{ks}^{(q+1)}$ coefficients in the expansion
of the transformation matrix.  These coefficients thus may enter
the transformed system as extra degrees of freedom, which we
will denote as the {\em parameters} of the transformed equation.
The canonical transformation can only be defined up to these
parameters. However in the context of a physical model, their
values will be constrained by the requirement that the solution
of the initial system belong to the Bargmann-Fock space. 

The remaining equations in the system (\ref{eqfa}) i.e. the
formulas corresponding to $l=q+2, q+3,...$ form a set of
recurrence equations for the $\{a_{rs}(z)\}$ matrix. This system
determines $\{a_{rs}(z)\}$ as a function of the parameters of the
transformed system. The procedure of determining the
transformation matrix for a given set of parameters is in
general infinite and it may be very difficult to find the
parameters that lead to solutions in the Bargmann-Fock space.
Hence it is conceivable that we know the initial (\ref{insy})
and the transformed (\ref{trsy}) systems of equations, without
being able to solve the transformation matrix (\ref{mata}). In
some cases this still allows us to draw some important
conclusions about features of physical interest, such as
degeneracies or symmetries. 

A special case arises if we assume that the expansion
(\ref{mata}) of $\{a_{rs}(z)\}$ in negative powers of $z$ is
finite. In this case confinement to the Bargmann-Fock space can
indeed easily be guaranteed. If the highest order in the
denominators of (\ref{mata}) is not greater than $\rho_{t}$ from
(\ref{fors}), which is the lowest power in the expansion of the
$F_{s}^{(t)}(z)$, then the corresponding solution of the initial
system is automatically analytic in the origin. In this case the
system (\ref{eqfa}) also remains finite and can be solved. This
procedure precisely leads to the isolated exact solutions of the
initial system.

       \section*{IV. The Solution of a Two Level System Coupled to a 
         Single Quantized Mode}
                \subsection*{A. The canonical form}

In this section we derive the canonical form of the dynamical 
equations for a two 
level system coupled to a single quantized mode. The 
Hamiltonian of such a system, sometimes called Rabi Hamiltonian 
\cite{reik}, is of the form
    \begin{equation}
     H = \omega a^{+}a + \mu \sigma_{3} + \lambda(\sigma^{+} + 
     \sigma^{-}) (a^{+} + a), \label{hami}
    \end{equation}
where $a^{+}$ and $a$ are boson field (\ref{koma}) creation and
annihilation operators, $\sigma^{\pm}=\frac{1}{2}(\sigma_{1} \pm i
\sigma_{2})$ and $\sigma_{1}$, $\sigma_{2}$, $\sigma_{3}$ are
Pauli matrices. The parameter $\omega$ is the boson field
frequency, $2\mu$ is the atomic level separation and $\lambda$
is the atom - boson field coupling constant. We choose the
energy unit in such a way that $\omega = 1$ and we assume that
$\lambda$ and $\mu$ are not vanishing simultaneously. 

The first step to solve the system (\ref{hami}) is to make a
unitary transformation which replaces operators $\sigma_{1}
\rightarrow \sigma_{3}$, $\sigma_{3} \rightarrow \sigma_{1}$ and
$\sigma_{2} \rightarrow - \sigma_{2}$. Then we write the
stationary Schr\"odinger equation for the two-component wave
function in the position variable $\left( \begin{array}{c}
f_{1}(\xi)\\f_{2}(\xi) \end{array} \right)$. In the second step,
by replacing $a^{+} \rightarrow z$ and $a \rightarrow
\frac{d}{dz}$, we perform a transition to a Bargmann-Fock space.
In this space the Schr\"odinger equation is equivalent to a
system of two first order differential equations for the
Bargmann-Fock space functions $f_{1}(z), f_{2}(z) \in {\cal F}
_{1}$
     \begin{eqnarray}
     \frac{d}{dz} f_{1}(z)&= &\frac{E-\lambda z}{z+\lambda} f_{1}(z) 
     - \frac{\mu}{z+\lambda} f_{2}(z) \nonumber \\ 
     \frac{d}{dz} f_{2}(z)&= &-\frac{\mu}{z-\lambda} f_{1}(z)  
     + \frac{E+\lambda z}{z-\lambda} f_{2}(z), \label{sysi}
     \end{eqnarray} 
where $E$ is an eigenenergy of ${\cal H}$. Note, that in the
Schr\"odinger representation of creation and annihilation
operators we have $a^{+} \rightarrow \frac{1}{\sqrt{2}}(\xi - i
p_{\xi})$, $a \rightarrow \frac{1}{\sqrt{2}}(\xi + i p_{\xi})$,
where $\xi$ and $p_{\xi}$ are conjugate position and momentum.
The system corresponding to (\ref{sysi}) in this representation
consist of two second order differential equations in a real
variable $\xi$. 
 
The present, Bargmann-Fock space formulation of the problem has
been investigated earlier and approximate solutions have been
found \cite{reik, schw}. In addition Ku\'s has derived 
some
isolated exact solutions, corresponding to degenerate levels
\cite{kus}. 

For our purposes we point out three properties of the system.
Firstly, it is of the form (\ref{insy}) with two finite
singularities in $z=\lambda, -\lambda$. Secondly, expanding
coefficients of (\ref{sysi}) in a Laurent series (\ref{expp}) we
find that $q=0$ and therefore the singular point at infinity is
of the first rank. Finally, we note the following symmetry: if
$\left( \begin{array}{c} f_{1}(z)\\f_{2}(z)\end{array} \right)$
is a solution of (\ref{sysi}) then $\left( \begin{array}{c}
f_{2}(-z)\\f_{1}(-z)\end{array} \right)$ is also a solution,
corresponding to the same energy value. 

The Birkhoff theorem is found to apply to the system in
(\ref{sysi}).  Hence one can claim the existence of a canonical
form. With $q=0$, the coefficients of this form will be
polynomials of the first degree! They can be obtained from
equations (\ref{eqfa}) and (\ref{zoep}), as explained in the
previous section. The linear terms of $P_{rs}(z)$ are inferred
at once from (\ref{zoep}) i.e. $P_{rs}^{(1)}(z)= (-1)^{r}
\lambda \delta_{rs}$. To calculate the remaining four zeroth
order terms of $P_{rs}(z)$ we use equation (\ref{eqfa}) with
$l=1$. The canonical form of the system (\ref{sysi}) is
therefore
     \begin{eqnarray}
     z\frac{d}{dz} F_{1}(z)&=& \left(E-\lambda z+\lambda^2\right) 
      F_{1}(z)
    + \left(-\mu - 2\lambda a_{12}^{(1)}\right) F_{2}(z) \nonumber \\
  z\frac{d}{dz} F_{2}(z)&=& \left(-\mu + 2\lambda a_{21}^{(1)}\right) 
     F_{1}(z) + \left(E+\lambda z+\lambda^2\right) F_{2}(z), 
     \label{syst}
     \end{eqnarray}
where $F_{1}(z)$ and $F_{2}(z)$ are linearly transformed 
$f_{1}(z)$ and $f_{2}(z)$ (\ref{tran}), and 
$a_{12}^{(1)}$, $a_{21}^{(1)}$ are parameters belonging to the 
transformation matrix (\ref{mata}).

The main feature of the canonical system is that it has only one 
singularity at $z=0$, and because of that is exactly solvable. Prior 
to solving it however we exploit the symmetry of solutions $\left(
\begin{array}{c} f_{1}(z)\\f_{2}(z)\end{array} \right)$ of the 
initial system (\ref{sysi}). To preserve this symmetry in the 
transformed pair, i.e. if $\left(\begin{array}{c} F_{1}(z)\\F_{2}(z)
\end{array} \right)$ is a solution of (\ref{syst}) then so is 
$\left(
\begin{array}{c} F_{2}(-z)\\F_{1}(-z)\end{array} \right)$, we 
must impose the following symmetry of the transformation matrix
     \begin{equation}
      a_{ij}(z) = a_{[i+1][j+1]}(-z), \hspace{1.5cm}i,j=1,2
       \label{syma}
     \end{equation}
where $[\cdot+\cdot]$ denotes an addition modulo $2$.

       \subsection*{B. Quantization conditions and canonical solutions}

Physical constrains require the solutions of (\ref{sysi}) to
belong to the Bargmann-Fock space, i.e. to converge in the
entire plane. This leads to quantization conditions, as we will
show in this section. 

The canonical system contains four parameters $\lambda$, $\mu$,
$E$ and $a_{12}^{(1)}$. In the usual perception of the problem
$\lambda$ and $\mu$ are external quantities, describing the
physics of the system, while $E$ and $a_{12}^{(1)}$ are
essentially free parameters, to be determined by the quantum
conditions. The indicial equation (\ref{indi}) corresponding to
the system (\ref{syst}) is seen to link $E$ and $a_{12}^{(1)}$: 
     \begin{equation}
      \rho = E+\lambda^{2} \pm A,
       \label{indk}
     \end{equation}
where $A=\mu + 2\lambda a_{12}^{(1)}$. For the system
(\ref{syst}) to have solutions in the Bargmann-Fock space at
least one value of $\rho$ must be a non-negative integer. The
value of $A$ thus determine the whole energy spectrum but in
turn it is controlled by the $a_{12}^{(1)}$ parameter, from which
the transformation matrix can be generated via
(\ref{eqfa}).  For the time being we will treat it as a free
parameter, delineating classes of solutions in the energy
spectrum. As we will see later its value will be fixed by the
requirement that the solutions of the original system
(\ref{sysi}) are in the Bargmann-Fock space. 

The general solution of (\ref{syst}) can be obtained by transforming 
it into a second order form
     \begin{equation}
      z^2 F_{1}''(z) + z\left[ 1-2\left(E+\lambda^2 \right)\right] 
      F_{1}'(z) + \left[\left(E+\lambda^2 \right)^2 -A^2 +\lambda 
      z - \lambda^2 z^2 \right] F_{1}(z) =0.    \label{sorf}
     \end{equation}
This equation 
can be reduced to a confluent hypergeometric (or Kummer) equation. 
Its general solution is a combination of two functions
     \begin{eqnarray}
      F_{1}(z)&=& C_{1}\:\exp(\lambda z) \: 
     _{1}F_{1}(1\!+\!A,1\!+\!2A;-2\lambda z)\: z^{E\!+\!\lambda^2 
      \!+\!A} + \nonumber \\ & & C_{2}\: \exp(\lambda z) \:
     _{1}F_{1}(1\!-\!A,1\!-\!2A;-2\lambda z)\: z^{E\!+\!\lambda^2 
      \!-\!A} .    \label{solu}
     \end{eqnarray}
with arbitrary $C_{1}$ and $C_{2}$. The function
$_{1}F_{1}(a,c;z)$ is called confluent series or Kummer function
and is defined for all complex $a$, $z$ and $c\neq -n$,
$n=0,1,2,...$
     \begin{equation}
      _{1}F_{1}(a,c;z)= 1+ \frac{a}{c} \frac{z}{1!}+ 
     \frac{a(a\!+\!1)}{c(c\!+\!1)}\frac{z^2}{2!}+           
     \frac{a(a\!+\!1)(a\!+\!2)}{c(c\!+\!1)(c\!+\!2)}\frac{z^3}{3!}
        +...\:.  \label{kumf}
     \end{equation}
The Kummer function is entire. The solutions for $F_{2}(z)$ are 
of the same general form (\ref{solu}), with however $z$ replaced 
by $-z$.

The asymptotic behavior of the solution (\ref{solu}) for
$|z|\rightarrow \infty$ is restricted by the function
$z^{\alpha_{1}}\exp(\lambda z) + z^{\alpha_{2}}\exp(-\lambda
z)$, where $\alpha_{1}$, $\alpha_{2}$ are real numbers
\cite{magn} and therefore the condition (\ref{waru}) is always
obeyed. Consequently the solution (\ref{solu}) belongs to the
Bargmann-Fock space ${\cal F}_{1}$ provided at least one of the
roots of the indicial equation (\ref{indk}) is a non-negative
integer. 

A particularly simple case occurs if $\lambda=0$ because then the
transformed system (\ref{syst}) coincides with the initial one
(\ref{sysi}). The transformation matrix reduces in this case to
the unit matrix. 

Let us now discuss the solutions (\ref{solu}) and their 
degeneracy for different values of $E+\lambda^2$. \\
(i) If $E\!+\!\lambda^2$ is not an integer nor a
half-integer then, to get physical solutions, we should take $A$
such that one of the numbers $\rho_{1}=E\!+\!\lambda^2\!+\!A$ and
$\rho_{2}=E\!+\!\lambda^2\!-\!A$ is a non-negative integer. In this 
case, according to the general rule $D_{1}$, one of the two functions
composing $F_{1}(z)$ can be entire. The corresponding solution
for $F_{2}(z)$ is also one-dimensional and is given by 
$F_{2}(z)=(-1)^{\rho_{t}+t} F_{1}(-z)$.\\
(ii) If $E\!+\!\lambda^2$ is a half-integer then we take $A$
also half-integral. Despite the fact that both
$E\!+\!\lambda^2\!+\!A$ and $E\!+\!\lambda^2\!-\!A$ are now
integers (rule $D_{2}$), The solutions $F_{1}(z)$ and $F_{2}(z)$
are still one-dimensional because in this case one of the
numbers $1\!+\!2A$ or $1\!-\!2A$ is a non-positive integer and
the corresponding Kummer function is not defined.\\
(iii) If $E\!+\!\lambda^2$ is an integer we can take the simple 
choice $A=0$. Then, for $E\!+\!\lambda^2 \geq 0$ each of the 
solutions is entire and forms a one-dimensional space (rule $D_{1}$),
but because the 
system (\ref{syst}) is diagonal we can always take its two linearly 
independent solutions of the form $\left(\begin{array}{c} F_{1}(z)\\
F_{2}(z)\end{array} \right)$ and $\left(\begin{array}{c} F_{1}(z)\\
-F_{2}(z)\end{array} \right)$. Consequently in this case the 
solutions are degenerate. This class of solutions will comprise 
the exact solutions found by Ku\'s.\\ 
(iv) If $E\!+\!\lambda^2$ is an integer and $E\!+\!\lambda^2
>0$, we can also take $A$ integral, $0<|A|\leq E\!+\!\lambda^2$. In
this case, although the system (\ref{syst}) is no more diagonal,
solutions $F_{1}(z)$ and $F_{2}(z)$ become two-dimensional and
lead to degenerate solutions. This represents the rarest case
of $D_{2}$ when two roots of the indicial equation are
integral and both corresponding solutions are analytical. 
 
To summarize we conclude that the only case where solutions are
degenerate can occur when $E+\lambda^2$ is a non-negative
integer.

     \subsection*{C. The isolated exact solutions}
 
In the previous section we have shown that the canonical form 
of the Rabi Hamiltonian can be solved within Bargmann-Fock 
conditions. This results in a coupling between $A$ and $E$ 
parameters, which is interesting in its own right, but does not 
lead to quantized energies. As we have already mentioned, the 
true quantization condition stems from the requirement, that 
the solutions of the original system belong to the 
Bargmann-Fock space. This implies that the transformation of 
the canonical solutions must act within the Bargmann-Fock 
space. In this way we fix $A$ and hence $E$.

In general this procedure is nontrivial since the 
transformation matrix is generated by an infinite system of 
equations (\ref{eqfa}), yielding an infinite series of 
coefficients $a_{rs}^{(k)}$. In this section we will not be 
concerned with the general case but only study the exactly 
solvable 
class of solutions, which corresponds to $A=0$. We assume that
the transformation is nontrivial i.e. $\lambda \neq 0$.

Let us first consider the simplest possibility $\mu=0$. With
$A=0$, and hence $a_{12}^{(1)}=0$, the indicial equation yields
that $\rho = E+\lambda^2$ must be a non-negative integer. It is
easy to find that the corresponding transformation matrix is
diagonal and reads
     \begin{equation}
      \left\{a_{rs}(z) \right\} = \left( \begin{array}{cc} 
      \left(1+\frac{\lambda}{z} \right)^{E+\lambda^{2}} & 0 \\ 
      0 &  \left(1-\frac{\lambda}{z} \right)^{E+\lambda^{2}}
      \end{array} \right), \label{maa1}
     \end{equation}
whereas the solutions are generated by
     \begin{eqnarray}
     F_{1}(z)&=&\exp(-\lambda z)\: z^{E+\lambda^2} \nonumber \\       
       F_{2}(z)&=&\exp(\lambda z)\: z^{E+\lambda^2}.
       \label{sla0}
     \end{eqnarray}
Note that in this case the expansion (\ref{mata}) in negative
powers of both diagonal terms terminate. The last non zero
coefficients are $a_{rs}^{(E+\lambda^{2})}$. On the other hand
the lowest power of $z$ in the expansion of the transformed
solution (\ref{solu}) is also $E+\lambda^{2}$. It is then easily
verified that the solution of the initial system, which follows
from (\ref{tran}), is of form (\ref{fodz}) and thus belongs
to the Bargmann-Fock space. The energy spectrum is obtained
directly from the indicial equation (\ref{indk}) and reads
$E_{\rho}=\rho - \lambda^2$, where $\rho= 0,1,2,...\:$. 

In the line of this example we now turn to the more general 
case $\mu \neq 0$ but keeping the $a_{rs}^{(k)}$ matrix finite. 
As a first example we assume that the expansion (\ref{mata}) of the
transformation matrix terminates, and $a_{rs}^{(1)}$ are the
only non-zero coefficients. In other words $a_{rs}^{(2)} =
a_{rs}^{(3)}=...=0$. In this case the system (\ref{eqfa}) is
also finite because starting from $l=3$ all the higher order
equations are equivalent to equations corresponding to $l=2$.
Taking into account symmetry properties (\ref{syma}) of the
matrix, the system reduces to four equations
     \begin{eqnarray}
      \overline{a}_{11} + 2\mu \overline{a}_{12} +2\lambda 
      \overline{a}_{12}^{2} - \lambda (E+\lambda^{2})&=&0 \nonumber\\
      2\mu \overline{a}_{11} + \overline{a}_{12} +2\lambda
      \overline{a}_{11} \overline{a}_{12} + \lambda\mu &=&0 
      \nonumber \\(E+\lambda^{2}) \overline{a}_{11} + \mu 
       \overline{a}_{12} - \lambda (E+\lambda^{2}) &=&0 \nonumber\\
       \mu \overline{a}_{11} + (E+\lambda^{2})\overline{a}_{12}
       +\lambda\mu &=&0    \label{sys1}
     \end{eqnarray}
where we denote $ \overline{a}_{kl}= a_{kl}^{(1)}$ for 
convenience. Let $\mu$ be non trivial and 
$0<\mu\leq(E+\lambda^{2})$. Then, after some algebra, one can show 
that the system (\ref{sys1}) is equivalent to the transformation 
matrix
     \begin{equation}
      \left\{a_{rs}(z) \right\} = \left( \begin{array}{cc}
      1+\frac{1\!+\!\mu^2}{4\lambda z} & -\frac{\mu}{2\lambda z} \\
      \frac{\mu}{2\lambda z} & 1-\frac{1\!+\!\mu^2}{4\lambda z}
      \end{array} \right) \label{maa2}
     \end{equation}
and two other conditions: $E+\lambda^{2}=1$ and $4\lambda^{2} + 
\mu^2=1$. From the $a_{12}(z)$ element in (\ref{maa2}) it 
also follows that $A=\mu+2\lambda a_{12}^{(1)}=0$.

Thus the very assumption that the transformation matrix terminates 
gave us the form of the matrix, the values of $E+\lambda^2$ and 
$A$, showing that they correspond to a degenerate solution of 
(\ref{solu}). In addition the system has energy $E= 1-\lambda^2$ 
only if the atomic level separation $2\mu$ and the atom-boson 
field coupling $\lambda$ obey condition $4\lambda^{2} +\mu^2=1$. 
The corresponding eigenfunction $\left(\begin{array}{c} 
F_{1}(z)\\F_{2}(z)\end{array} \right)$ is given by (\ref{sla0}) and  
$\left(\begin{array}{c} f_{1}(z)\\f_{2}(z)\end{array} \right)$can be 
found by 
applying (\ref{tran}). One can easily shown that $f_{1}(z)$ and 
$f_{2}(z)$ belong to the Bargmann-Fock space. This 
solution agrees with the first root of the Ku\'s series \cite{kus}.

In an analogous way one can show that if the transformation matrix 
terminates from the third order coefficients 
onwards $a_{rs}^{(3)} = 
a_{rs}^{(4)}=...=0$ and $0<\mu\leq(E+\lambda^{2})$ then the 
transformation matrix functions are
     \begin{eqnarray}  
      a_{11}(z)&=\hspace{0.3cm} a_{22}(-z)&=\hspace{0.3cm}  
      1+\frac{2\lambda\!+\!\frac{\mu^2}{2\lambda}}{z}+\frac
     {\lambda^2\!+\!\mu^2\!+\!\frac{\mu^4-\mu^2}{8\lambda^2}}{z^2}
     \nonumber\\
      a_{12}(z)&=\hspace{0.3cm} a_{21}(-z)&=\hspace{0.3cm}
      -\frac{\mu}{2\lambda z} + \frac{\mu (6\lambda^2 +\mu^2-1)}
      {4\lambda^2 z^2}     \label{maa3}
     \end{eqnarray}
i.e. again $A=0$ and the remaining conditions are $E+\lambda^2 = 2$ 
and $32\lambda^4 -32\lambda^2+12\lambda^2 \mu^2 -5\mu^2 +\mu^4 +4 
=0$, which corresponds to the second root of the Ku\'s series. 
In this way all solutions of the Rabi Hamiltonian characterized by a 
terminating $\left\{a_{rs}(z) \right\}$ can be generated. They are found to 
coincide with all known exact solutions.

\section*{V. Discussion}

The Rabi Hamiltonian describes the coupling between a two level 
system and a single mode through a linear interaction term $ 
\lambda \sigma_{1} (a^{+} + a)$. This problem is relevant in 
quantum optics but also appears in molecular physics as the 
simplest example of vibronic coupling \cite{fult, brin}.

In Bargmann-Fock space it can be represented by a system of two 
first order differential equations, as shown in (\ref{sysi}). 
For $\mu=0$ this system separates into two independent equations
     \begin{eqnarray}
     (z+\lambda) \frac{d}{dz} f_{1}(z)&= &(E-\lambda z) 
     f_{1}(z)  \nonumber \\
     (z-\lambda) \frac{d}{dz} f_{2}(z)&= &(E+\lambda z)         
     f_{2}(z). \label{smd1}
     \end{eqnarray}
This situation corresponds to a highly symmetric case with two 
uncoupled harmonic oscillators that are displaced to the left 
and to the right in coordinate space \cite{schw}, and cross at 
the coordinate origin. From this perspective the introduction 
of the level separation $2\mu$ may be viewed as a symmetry 
lowering perturbation, which couples the two oscillators. 

If we 
now compare the original system (\ref{sysi}) to the canonical 
one (\ref{syst}), we still have essentially a set of two 
displaced oscillators - but now the coupling term no longer 
corresponds to $\mu$ but to  $\mu + 2\lambda a_{12}^{(1)}$! 
Hence the canonical transformation provides a degree of freedom 
that allows to perform a displacement in the space of the 
coupling parameter itself. We can thus adjust $a_{12}^{(1)}$ in 
such a way that it compensates for the energy gap, by 
requiring  $a_{12}^{(1)}=\frac{-\mu}{2\lambda}$, or $A=0$. 
System (\ref{syst}) then becomes
     \begin{eqnarray}
     z \frac{d}{dz} F_{1}(z)&= &(E-\lambda z+\lambda^2)
     F_{1}(z)  \nonumber \\
     z \frac{d}{dz} F_{2}(z)&= &(E+\lambda z+\lambda^2)
     F_{2}(z). \label{smd2}
     \end{eqnarray}
This system again describes two degenerate harmonic oscillators 
that are displaced in coordinate space and also underwent a 
translation in $z$ space over $+$ or $-\lambda$.

We thus have shown that the exact solutions , that are found 
for $A=0$, can be mapped onto the energy spectrum of a 
degenerate harmonic oscillator. A similar result can also be 
obtained for the Juddian exact solutions of the $E\otimes e$ 
Jahn-Teller Hamiltonian \cite{szop}.

\section*{Acknowledgments}

We are indebted to the Belgian Government (Ministerie voor de 
Programmatie van het Wetenschapsbeleid) for financial support. 
M.S. thanks the Commission of the European Union for a 
mobility grant and the Polish KBN for grant PB 1108/P03/95/08.

\end{document}